\documentstyle[12pt,openbib]{article}
\newcommand{\R}{I\!\! R}
\newcommand{\h}{\widehat{\:}}

\title{Quantum Field Theory in Curved Spacetime}
\author{Robert M. Wald\\
         {\it Enrico Fermi Institute and Department of Physics}\\
         {\it University of Chicago}\\
         {\it 5640 S. Ellis Avenue}\\
         {\it Chicago, Illinois 60637-1433}}
\date{}

\begin{document}
\maketitle
\begin{abstract}

We review the mathematically rigorous formulation of the quantum
theory of a linear field propagating in a globally hyperbolic
spacetime. This formulation is accomplished via the algebraic
approach, which, in essence, simultaneously admits all states in all
possible (unitarily inequivalent) Hilbert space constructions.  The
physically nonsingular states are restricted by the requirement that
their two-point function satisfy the Hadamard condition, which insures
that the ultra-violet behavior of the state be similar to that of the
vacuum state in Minkowski spacetime, and that the expected
stress-energy tensor in the state be finite. We briefly review the
Unruh and Hawking effects from the perspective of the theoretical
framework adopted here. A brief discussion also is given of several
open issues and questions in quantum field theory in curved spacetime
regarding the treatment of ``back-reaction", the validity of some
version of the ``averaged null energy condition'', and the formulation
and properties of quantum field theory in causality violating
spacetimes.

\end{abstract}
\newpage

\section{Introduction}
\label{intro}

The subject of quantum field theory in curved spacetime is the study of
the behavior of quantum fields propagating in a classical gravitational
field. It is used to analyze phenomena where the quantum nature of fields
and the effects of gravitation are both important, but where the quantum
nature of gravity itself is assumed not to play a crucial role, so that
gravitation can be described by a classical, curved spacetime, as in the
framework of general relativity.

The main initial development of the theory occurred in the late 1960's,
driven primarily by the desire to analyze the
phenomenon of particle creation occurring in the very early universe.
By 1969, one can find the theory formulated in recognizably modern
form and applied to cosmology
in the paper of Parker \cite{par}. In the early 1970's, the theory was
applied to the study of particle
creation near rotating and charged black
holes, where the discovery of classical ``superradiant scattering" (analogous
to stimulated emission) strongly suggested that spontaneous particle
creation should occur. This line of research culminated in the analysis by
Hawking of particle creation resulting from the gravitational collapse
of body to form a black hole \cite{haw}. It thereby was discovered that
black holes radiate as perfect black bodies at temperature
$T = \kappa/2 \pi$, where $\kappa$ denotes the surface gravity of the
black hole. This result solidified an undoubtedly deep relationship between
the laws of black hole physics and the laws of thermodynamics, the
ramifications of which continue to be pondered today.

As a direct consequence of Hawking's remarkable discovery, there
occurred in the mid-to-late 1970's a rapid and extensive development
of the theory of quantum fields in curved spacetime and its applications
to a variety of phenomena. A good summary of this body of work can be
found in the monograph of Birrell and Davies \cite{bd}. Further
important applications to cosmology were made in the early 1980's,
as the methods and results of quantum field theory in curved
spacetime were used to calculate the perturbations generated by
quantum field fluctuations during inflation. Many of these lines of
investigation begun in the late 1970's and early 1980's continue to
be pursued today.

Although it would be more difficult to point to major historic
landmarks, during the past twenty
years the theoretical framework of quantum field theory in curved
spacetime has undergone significant development, mainly through the
incorporation of key aspects of the algebraic approach to quantum field
theory. As a result, the theory of a linear quantum field propagating in a
globally hyperbolic spacetime can be formulated in an entirely
mathematically rigorous manner insofar as the definition of the
fundamental field observables is concerned.

My main goal here is to review the key developments leading to a
mathematically rigorous formulation
of quantum field theory in curved spacetime. (Much more detail can be
found in my recent book on this subject \cite{w1}, to which I refer the
reader for a more comprehensive and pedagogically oriented discussion.) I
will also briefly review some open issues and questions regarding
the treatment of ``back-reaction", positivity properties of the
expected stress-energy tensor, and the formulation
and properties of
quantum field theory in causality violating spacetimes. Notational
conventions follow those of \cite{w2}.

\section{The Formulation of Quantum Field Theory in Curved Spacetime}
\label{form}

In the classical mechanics of a system with $n$
degrees of freedom, the state of a system at any instant of time is
described by a point in phase space,
${\cal M}$, which has the structure of a $2n$-dimensional
symplectic manifold, i.e., on ${\cal M}$
is defined a non-degenerate, closed two-form $\Omega_{ab}$,
referred to as a symplectic form.
Most commonly, ${\cal M}$ is obtained as the cotangent bundle of an
$n$-dimensional configuration manifold ${\cal Q}$, in which case
$\Omega_{ab}$ is given by
\begin{equation}
\Omega_{ab} = \sum_{\mu = 1}^{n} 2 \nabla_{[a} p_{|\mu|} \nabla_{b]} q^\mu
\label{Om}
\end{equation}
where $(q^\mu, p_\mu)$ denote (local) canonical coordinates on the
cotangent bundle. (The $2$-form $\Omega$ is then independent of the
choice of these coordinates.) An observable
in classical mechanics is simply a real-valued function on ${\cal M}$.

This basic structure of the classical description of a system stands in
marked contrast to the corresponding quantum description. In quantum
mechanics, the state of a system at a given instant of time is described
by a vector (or, more generally, a density matrix) in an
infinite-dimensional, separable Hilbert space ${\cal F}$. An observable
is a self-adjoint operator on ${\cal F}$. Since all
infinite-dimensional, separable Hilbert spaces are isomorphic to each
other, the content of a quantum theory corresponding to a given
classical theory is completely specified by giving a map
$\h : {\cal O}_c \rightarrow {\cal O}_q$, where ${\cal O}_c$ denotes the
set of classical observables (i.e., real valued functions on ${\cal M}$), and
${\cal O}_q$ denotes the set of quantum observables
(i.e., self-adjoint operators on ${\cal F}$).

Since the structures
of the classical and quantum theories are so different, it is far from
obvious how the map $\h$ is to be determined. However, a key
guiding principle arises from the comparison of the classical and
quantum dynamics. In classical mechanics with Hamiltonian $H$, the
rate of change of an observable, $f$, in the ``Heisenberg representation"
is given by
\begin{equation}
df/dt = \{f, H \}
\label{ceom}
\end{equation}
where $\{ , \}$ denotes the Poisson bracket, defined by
\begin{equation}
\{f, g \} = \Omega^{ab} \nabla_a f \nabla_b g
\label{pb}
\end{equation}
where $\Omega^{ab}$ is the inverse of $\Omega_{ab}$. On the other hand,
in quantum mechanics with Hamiltonian $\hat{H}$, the
rate of change of an observable, $\hat{f}$, in the Heisenberg
representation is given by
\begin{equation}
d\hat{f}/dt = - i [\hat{f}, \hat{H}]
\label{qeom}
\end{equation}
Consequently, there will be a close correspondence between classical
and quantum dynamics (for any choice of Hamiltonian) if the map,
$\h$ , can be chosen so as to satisfy the ``Poisson bracket goes to
commutator" rule:
\begin{equation}
[\hat{f}, \hat{g}] =  i \widehat{\{f, g \}}
\label{pbc}
\end{equation}

In fact, it is well known that even in standard Schrodinger quantum
mechanics, no map $\h$ exists which implements the relation
(\ref{pbc}) on all observables (see, e.g., \cite{che}). However, the relation
(\ref{pbc}) can be implemented on a restricted class of observables. In
particular, consider the case where ${\cal M}$ has the structure of a
symplectic vector space, i.e., ${\cal M}$ is a vector space and the
symplectic form
$\Omega_{ab}$ has constant components in a globally parallel basis,
so that it may be viewed as an antisymmetric, bilinear map
$\Omega: {\cal M} \times {\cal M} \rightarrow \R$ rather than a tensor
field on ${\cal M}$. (This situation arises whenever ${\cal M}$ is the
cotangent bundle of a configuration space ${\cal Q}$
which has vector space structure and $\Omega$
is defined by eq.(\ref{Om}).) In this case, we will refer to the
$2n$-dimensional vector space of linear
functions on ${\cal M}$ as the {\it fundamental classical observables},
since any classical observable (i.e., function on ${\cal M}$) can be
expressed as a function of the $2n$ elements of a basis for the linear
observables. If we restrict attention to the fundamental observables, then
the Poisson-bracket-commutator relationship (\ref{pbc}) {\it can} be
implememented. Moreover, in the sense explained below,
it {\it uniquely} determines a map $\h$ on these observables.

Before stating this result more precisely, it is useful
to note that since $\Omega$ is non-degenerate, any fundamental
observable (i.e., linear function) on ${\cal M}$ can be written in the form
$\Omega (y, \cdot)$ for some $y \in {\cal M}$, where by
$\Omega (y, \cdot)$ we mean the function $f: {\cal M} \rightarrow \R$
defined by $f(z) = \Omega (y, z)$. In this notation,
the Poisson bracket of the fundamental observables is given by
\begin{equation}
\{\Omega (y_1, \cdot), \Omega (y_2, \cdot) \} =
- \Omega (y_1, y_2) 1
\label{pb2}
\end{equation}
where $1$ denotes the function on ${\cal M}$ which takes the value $1$
at each point.
Hence, the Poisson-bracket-commutator
relationship (\ref{pbc}) for the fundamental
observables is simply
\begin{equation}
[\hat{\Omega} (y_1, \cdot), \hat{\Omega} (y_2, \cdot)] =
- i \Omega (y_1, y_2) I
\label{pbc2}
\end{equation}
where $I$ denotes the identity operator on the
Hilbert space.

However, there are some potential
technical difficulties with eq.(\ref{pbc2}) because
$\hat{\Omega} (y, \cdot)$ should be an unbounded operator and, hence,
can be defined only on a dense domain, so compositions (and, hence,
commutators) are not automatically well
defined. These difficulties are most easily
dealt with by working with the following exponentiated version
of (\ref{pbc2}): For each $y$, define the classical observable $W(y)$ by
\begin{equation}
W(y) = {\rm exp}[- i \Omega (y, \cdot)]
\label{W}
\end{equation}
Then eq.(\ref{pbc2}) together with the self-adjointness of
$\hat{\Omega} (y, \cdot)$ is formally equivalent to the following
{\it Weyl relations}
\begin{equation}
\hat{W}(y_1) \hat{W}(y_2) =
{\rm exp}[ i \Omega (y_1, y_2)/2] \hat{W}(y_1+y_2)
\label{Weyl1}
\end{equation}
\begin{equation}
\hat{W}^{\dag} (y) = \hat{W}(-y)
\label{Weyl2}
\end{equation}
We shall view the Weyl relations as providing a mathematically
precise statement of the Poisson-bracket-commutator relationship,
which avoids operator domain problems.

The key uniqueness result is provided by the Stone-von Neumann
theorem: {\it For a finite dimensional symplectic vector space
$({\cal M}, \Omega)$, any two strongly continuous (in $y$), irreducible
representations of the Weyl relations (\ref{Weyl1}), (\ref{Weyl2})
are unitarily equivalent.} Thus, for a classical linear system with finitely
many degrees of freedom, the ``Poisson bracket goes to commutator"
rule determines in a natural, canonical way a corresponding quantum
theory in so far as the fundamental observables are concerned.
This provides the main justification for the standard choices of the
Cartesian position and momentum operators for Schrodinger
quantum mechanics, as found in standard texts.
Note, however, that although
all classical observables can be written as functions of the
fundamental observables, ``factor ordering" ambiguities
generally arise when
one attempts to express an arbitrary quantum observable as a function
of the quantum representatives of the fundamental observables. Since
the ``Poisson bracket goes to commutator" rule cannot be implemented
for all observables, there does not appear to be any natural way to
resolve this factor ordering ambiguity. Thus,
when a sufficiently general class of observables is considered, it appears
that there are many quantum theories corresponding to a given classical
theory. However, our primary interest here is in the fundamental
observables, where the Stone-von Neumann theorem does provide the
desired uniqueness result for a system with finitely many degrees of
freedom.

All of the above basic structure present
in a classical system with finitely many
degrees of freedom also is present in the theory of a classical field
propagating in a globally hyperbolic spacetime
$(M, g_{ab})$. Consider, for definiteness,
a linear, Klein-Gordon scalar field, $\phi$, satisfying
\begin{equation}
\nabla^a \nabla_a \phi - m^2 \phi = 0
\label{KG}
\end{equation}
This equation has a well posed initial value formulation, with the initial
data consisting of the pair of functions $(\phi, \pi)$ on a Cauchy surface
$\Sigma$, where $\pi = n^a \nabla_a \phi$, with $n^a$ the unit normal
to $\Sigma$. We have a choice of precisely what class of functions to
allow in phase space ${\cal M}$, but a particularly convenient choice
is to require $\phi$ and $\pi$ to be smooth and of compact support, i.e.,
we define
\begin{equation}
{\cal M} = \{ (\phi, \pi) | \phi, \pi \in C^\infty_0 (\Sigma) \}
\label{M}
\end{equation}
The Klein-Gordon Lagrangian gives rise to
a well defined, conserved symplectic structure
$\Omega: {\cal M} \times {\cal M} \rightarrow \R$ on ${\cal M}$,
given by
\begin{equation}
\Omega [(\phi_1, \pi_1), (\phi_2, \pi_2)] =
\int_\Sigma (\pi_1 \phi_2 - \pi_2 \phi_1)
\label{OmKG}
\end{equation}

We also have a choice as to precisely which linear observables on
${\cal M}$ should be viewed as the ``fundamental observables". A natural
choice is to take the fundamental observables to consist of all linear maps
from ${\cal M}$ into $\R$ which are of the form $\Omega (y, \cdot)$ with
$y \in {\cal M}$. On this class of observables, there is a well defined
Poisson bracket, given again precisely by eq.(\ref{pb2}).

It is worth pointing out that for any ``test function", $f$, on spacetime
(i.e., any $f \in C^\infty_0 (M)$) and any solution, $\psi$, of the
Klein-Gordon equation with initial data of compact support
on a Cauchy surface, we have the identity (see lemma 3.2.1 of \cite{w1})
\begin{equation}
\int_M f \psi = \Omega (Ef, \psi)
\label{id}
\end{equation}
where $Ef$ denotes the advanced minus retarded solution with source $f$,
and the natural identification of phase space with the space of solutions
to the Klein-Gordon equation is implicit on the right side of this equation.
Consequently, the fundamental observable $\Omega (y, \cdot)$ on phase
space also may be viewed as the observable on solution space
defined by the spacetime smearing of solutions with a particular
test function.
The corresponding quantum observable $\hat{\Omega} (y, \cdot)$ will
then have an alternative interpretation of being a spacetime smearing
of the Heisenberg field operator. Our choice of fundamental observables
thus corresponds to the collection of Heisenberg field operators smeared
with arbitrary test functions $f$.

We have seen above that the theory of a linear Klein-Gordon field on
a globally hyperbolic spacetime has a phase space structure and a choice
of fundamental observables that parallels completely the case of
a system with finitely many degrees of
freedom. Thus, the only relevant difference which occurs when we
attempt to construct a quantum theory of a field
arises from the fact that now
dim$({\cal M}) = \infty$. However, this difference is crucial because
the Stone-von Neumann theorem does not hold in infinite dimensions, and
there are many inequivalent ways of implementing the ``Poisson bracket
goes to commutator" rule for the fundamental observables.

The class of possible quantum field theory constructions can be
restricted in a natural way by exploiting the
analogy of a quantum field with
an infinite collection of harmonic oscillators, and performing a construction
analogous to the standard construction of the quantum theory of a
harmonic oscillator using annihilation and creation operators. A
mathematically precise implementation of this idea can be achieved
by introducing a real inner product $2 \mu$
on ${\cal M}$ which makes $\Omega$ be norm preserving, i.e., we
introduce a symmetric, bilinear map
$\mu: {\cal M} \times {\cal M} \rightarrow \R$ such that
\begin{equation}
\mu (\psi_1, \psi_1) = \frac{1}{4} \sup_{\psi_2 \neq 0}
\frac{|\Omega(\psi_1, \psi_2)|^2}{\mu (\psi_2, \psi_2)}
\label{mu}
\end{equation}
It can be shown (see \cite{w1}) that there always exist a wide class
of $\mu$'s satisfying (\ref{mu}). Given a $\mu$ satisfying (\ref{mu}),
we complete ${\cal M}$ in the inner product $2 \mu$. We then use
$\Omega$ to define a complex structure, so as to convert the completion
of ${\cal M}$ into a complex Hilbert space ${\cal H}$. We
construct the symmetric Fock space ${\cal F}_S ({\cal H})$
based upon ${\cal H}$ by
\begin{equation}
{\cal F}_S ({\cal H}) = C \oplus {\cal H} \oplus ({\cal H} \otimes_S {\cal H})
\oplus ...
\label{Fock}
\end{equation}
where $C$ denotes the complex numbers and
${\cal H} \otimes_S {\cal H}$ denotes the symmetrized tensor
product of ${\cal H}$ with itself.
Finally, we define the fundamental quantum observables
$\hat{\Omega} (y, \cdot)$ in terms of annihilation and creation operators
on ${\cal F}_S ({\cal H})$, in close analogy with the harmonic oscillator
construction. Details of this construction can be found in \cite{w1}.

However, if two maps $\mu_1$ and $\mu_2$ satisfying eq.(\ref{mu})
are sufficiently different in the precise sense stated in theorem 4.4.1
of \cite{w1}, the quantum field constructions based upon them will
be unitarily inequivalent. Thus, we must still face the key question:
Which $\mu$ should we choose to do the quantum field construction?

In Minkowski spacetime -- and, more generally, in curved, stationary
spacetimes -- the presence of a time translation symmetry gives rise to
the following natural, ``preferred" choice of $\mu$, which is motivated
by the choice of $\mu$ which simplifies the description of the time
independent, quantum mechanical harmonic oscillator: For
$y_1, y_2 \in {\cal M}$, we define $\mu(y_1, y_2)$ to be the real part
of the Klein-Gordon inner product of the positive frequency parts of
the solutions corresponding to the initial data $y_1$ and $y_2$. (A more
mathematically precise and complete description of this construction
can be found in section 4.3 of
\cite{w1}.) Furthermore, for this construction in stationary
spacetimes, states in ${\cal F}_S ({\cal H})$ then have a natural,
physical interpretation
in terms of ``particles", with the one-dimensional subspace $C$
corresponding to the ``vacuum", with ${\cal H}$ corresponding to the
subspace of single particle states, with
${\cal H} \otimes_S {\cal H}$ corresponding to the subspace
of two-particle states, etc.

However, in a general, non-stationary spacetime, there does not appear to
any mathematically preferred choice of $\mu$ nor any physically
preferred definition of ``particles" -- although the Hadamard condition
(discussed below) does pick out a preferred unitary equivalence class of
$\mu$'s for spacetimes with compact Cauchy surfaces \cite{w1}.
So the question remains: Which $\mu$
should one choose? Equivalently, how should one define the notion
of ``particles" in a general, curved spacetime?

My view is that this question has roughly the same status in
quantum field theory as the following question in
classical general relativity: Which coordinate system should one choose in
a general, curved spacetime? This latter question is very natural
one to ask (and often is asked by begining students) if one has
learned special relativity via its formulation in
terms of global inertial coordinate systems. It is ``answered" by
formulating
general relativity in a geometrical, coordinate independent manner, so that
the question becomes manifestly irrelevant. Similarly, the issue of how
to define ``particles" is a natural one to ask by those to whom quantum
field theory was presented as though it were a theory of ``particles". It also
can be answered by reformulating the theory in a manner which makes
the question manifestly irrelevant.

In order to do this, we need a framework for quantum field theory which
(initially, at least) simultaneously admits all states occurring in all
(unitarily inequivalent) Hilbert space constructions of the theory, so that
no ``preferred construction" need be specified in advance.
The algebraic approach
accomplishes this goal in a mathematically elegant and straightforward
manner, and I shall now briefly describe some of the main elements of the
algebraic formulation.

The key idea of the algebraic approach is to reverse the logical order in
which the notion of states and observables are specified. One first
introduces a notion of {\it observables} and provides the set of
observables with the
mathematical structure of an abstract *-algebra, ${\cal A}$. (For some
purposes, it is required that ${\cal A}$ have the additional structure of
being a C*-algebra.) On then defines a {\it state} to be a linear map
$\omega: {\cal A} \rightarrow C$ which satisfies the positivity condition
\begin{equation}
\omega (A^*A) \geq 0
\label{pos}
\end{equation}
for all $A \in {\cal A}$, as well as the normalization condition
$\omega (I) = 1$, where $I$ denotes the identity element of the
algebra. Although this notion of a state may appear to be drastically
different from the usual notion of a state as a vector (or, more
generally, a density matrix) in a Hilbert space, there is, in fact, a very
close relationship between these notions: If ${\cal F}$ is a Hilbert space
and ${\cal A}$ is a sub-algebra of the C*-algebra, ${\cal L}({\cal F})$,
of bounded linear maps on ${\cal F}$, then any density matrix state
$\rho$ on ${\cal F}$ gives rise to an algebraic state, $\omega$,
on ${\cal A}$ by the formula
$\omega (A) = {\rm tr} (\rho A)$
for all $A \in {\cal A}$. Conversely, if $\omega$ is an algebraic state on
the C*-algebra, ${\cal A}$, then the GNS construction shows that there
exists a Hilbert space ${\cal F}$, a representation
$\pi: {\cal A} \rightarrow {\cal L}({\cal F})$, and a cyclic vector
$\Psi \in {\cal F}$ such that for all $A \in {\cal A}$ we have
$\omega (A) = <\Psi |\pi (A) |\Psi>$. Thus, every state in the algebraic
sense corresponds to a state in the usual sense in some Hilbert space
construction of the quantum field theory. The key advantage of the
algebraic approach is that it allows one to consider, on an equal footing,
all states arising in all unitarily inequivalent Hilbert space constructions
of the theory.

To define the theory of a quantum field in a curved,
globally hyperbolic spacetime
via the algebraic approach, we must specify a C*-algebra
structure on the class of observables that we wish to consider. If one
wishes -- initially, at least -- merely to have the fundamental
observables defined in the theory, then a
C*-algebra structure can be specified
as follows: We start with the classical phase space ${\cal M}$, eq.(\ref{M}),
with symplectic structure $\Omega$ given by eq.(\ref{OmKG}). We then
choose an inner product $\mu$ on ${\cal M}$
satisfying eq.(\ref{mu}), and we perform the Hilbert space
quantum field theory construction outlined above to define
the observables $\hat{\Omega} (y, \cdot)$ as self-adjoint operators on
a Fock space ${\cal F}$. Next, we define corresponding unitary operators
$\hat{W} (y)$ by exponentiation (see eq.(\ref{W})). These operators
satisfy the Weyl relations (\ref{Weyl1}), (\ref{Weyl2}). The finite linear
combinations of the $\hat{W} (y)$'s then have the natural structure of a
*-algebra. Completion of this algebra in the norm provided by
${\cal L}({\cal F})$ defines the desired C*-algebra, ${\cal A}$. In principle,
${\cal A}$ could depend upon the choice of $\mu$. If it did, then we would
be back in a situation very similar to the situation we faced when we had
many unitarily inequivalent Hilbert space constructions of the theory;
we now would have many different possible constructions of the
algebra of fundamental observables and
no obvious means of choosing a ``preferred" one. Fortunately, this is not
the case: Even when the inner products $\mu_1$ and $\mu_2$ yield
unitarily inequivalent Hilbert space constructions of a quantum field
theory, the algebras ${\cal A}_1$, ${\cal A}_2$ to which they give rise are
isomorphic as abstract
C*-algebras. In other words, associated with the symplectic vector
space $({\cal M}, \Omega)$ for a Klein-Gordon (or other linear,
bosonic) field
on a given globally hyperbolic spacetime is a unique, well defined
C*-algebra of fundamental observables -- known as
the {\it Weyl algebra} -- constructed in the manner described above,
using any choice of $\mu$ satisfying (\ref{mu}).

The specification of the Weyl algebra, ${\cal A}$, as the C*-algebra of
observables
completes the formulation of the quantum theory of a Klein-Gordon
field in an arbitrary globally hyperbolic, curved spacetime insofar as
the definition of fundamental observables is concerned. Note that
the specification of a state on ${\cal A}$ corresponds, roughly, to the
specification of the complete list of all $n$-point distributions
$<\phi(x_1),...,\phi(x_n)>$, subject to all of the conditions arising from
the positivity requirement (\ref{pos}).

However, there are other observables besides the fundamental
observables which one may wish to consider. Most
prominent among these is the stress-energy tensor,
$T_{ab}$, of the quantum field,
since it is needed to describe ``back-reaction" effects of the quantum
field on the gravitational field.
In order to extend the construction of quantum field theory in curved
spacetime to encompass the stress-energy tensor, one would like to
enlarge the Weyl algebra ${\cal A}$ to a new algebra of observables
which contains elements corresponding to the
(presumably, spacetime smeared) stress-energy tensor $T_{ab}$.
A first step in this regard would be to define ${\hat{T}}_{ab}$ as
an operator-valued-distribution in Hilbert space constructions of the
theory. There are serious difficulties with doing this because, formally,
${\hat{T}}_{ab}$ is the product of two distributions at the same spacetime
point, so some ``regularization" is needed to give it a mathematically well
defined meaning. Some progress towards defining ${\hat{T}}_{ab}$ as
an operator-valued-distribution has been reported recently \cite{koh},
but most work to date has focused on the less ambitious goal of defining
expectation values of the (unsmeared)
stress-energy tensor. (In a Hilbert space construction, this corresponds to
defining the stress-energy tensor as a quadratic form
on the Hilbert space rather than as an operator-valued-distribution.)
Note that for an algebraic state $\omega$, a knowledge of
$<T_{ab}>_\omega$ is precisely what is needed in order to determine
whether the semiclassical Einstein equation
\begin{equation}
G_{ab} = 8 \pi <T_{ab}>_\omega
\label{see}
\end{equation}
is satisfied, so a knowledge of $<T_{ab}>$ is all that is needed to analyze
back reaction within the context of the semiclassical approximation.

The main results of the analysis of $<T_{ab}>$ are the following
(see \cite{w1} for more details): (i) $<T_{ab}>_\omega$ can be defined only
for states, $\omega$, that satisfy the
{\it Hadamard condition}, which, in essence, states that the ``ultra-violet"
behavior of the state -- as measured by the short distance behavior of
the two point distribution $<\phi(x) \phi(x')>_\omega$ -- is similar
in nature to the
short distance behavior of the two-point distribution for the vacuum
state in Minkowski spacetime. (A precise definition of the ``global
Hadamard condition" can be found in \cite{kw}; its equivalence to a
``local Hadamard condition" was proven in \cite{r}.) States which fail
to satisfy the Hadamard condition are to be viewed as ``physically
singular", in that their stress-energy is infinite (or otherwise ill defined).
The Hadamard condition thus provides an important
additional restriction on the
class of states which otherwise would be admissible when only the
fundamental observables are considered.
(ii) The prescription for assigning an expected stress-energy,
$<T_{ab}>_\omega$, to all Hadamard states, $\omega$, is uniquely
determined up to addition of (state independent) conserved local
curvature terms, by a list of physical
properties that $<T_{ab}>_\omega$ should
satisfy. The ``point-splitting" regularization prescription satisfies these
properties and thus provides a completely satisfactory definition of
$<T_{ab}>_\omega$, up to the local curvature ambiguity.

Thus, the status of quantum field theory for a linear field in a globally
hyperbolic spacetime, $(M, g_{ab})$, may be summarized as follows:
{}From $(M, g_{ab})$ and the classical symplectic structure of the field,
we can construct the Weyl algebra, ${\cal A}$, of fundamental
observables (or a corresponding ``anticommutator algebra" in the case
of fermion fields). States are then defined in the algebraic sense with
respect to ${\cal A}$. Nonsingular states must, in addition,
satisfy the Hadamard condition. For Hadamard states, $<T_{ab}>$ is
well defined up to the ambiguity of adding conserved local curvature
terms.

We conclude this section by briefly describing the statements of the
Unruh and Hawking effects within the framework developed above.
(Again, much more detail can be found in \cite{w1}.) In the Unruh effect,
one considers the action of a one-parameter family of Lorentz boosts
on Minkowski spacetime. For definiteness, we normalize the boost Killing
field to have unit norm on an orbit of acceleration $a$.
We note that the orbits of the boost isometries
are timelike in the ``right wedge" (as well as the ``left wedge") region of
Minkowski spacetime. Furthermore, this ``right wedge" region -- when
viewed
as a spacetime in its own right -- is globally hyperbolic. Hence, there is a
well defined Weyl algebra, ${\cal A}_R$, of ``right wedge" observables,
which is naturally a subalgebra of the Weyl
algebra, ${\cal A}$, for all of Minkowski spacetime. Hence, the restriction
of the ordinary Minkowski vacuum state, $\omega_0$, to the subalgebra
${\cal A}_R$ defines a state on the right wedge spacetime. The Unruh
effect is the assertion that this state is a thermal state at temperature
$T = a/2 \pi$, in the precise sense that it satisfies the KMS condition
with respect to the notion of ``time translations" provided by
the Lorentz boosts. In fact, this mathematical statement of the Unruh
effect was proven by Bisognano and Wichmann \cite{bw}, independently
of (and simultaneously with) the paper of Unruh \cite{u}. However, the
physical interpretation of this fact -- namely, that an observer with
acceleration $a$
will ``feel himself" to be immersed in a thermal bath at temperature
$T = a/2 \pi$ when the field is in the Minkowski vacuum state -- is due
to Unruh.

In the Hawking effect, one considers a spacetime describing the
gravitational collapse of a body to form a black hole. One assumes that
the black hole settles down to a stationary final state, with constant
surface gravity $\kappa$. The Hawking effect \cite{haw}, \cite{fh}
is the assertion that any nonsingular (i.e., Hadamard) state
asymptotically approaches, at late times, a thermal state at temperature
$T = \kappa/2 \pi$ with respect to the subalgebra associated with
solutions which ``appear to emerge from the direction of the black hole".
In other words, a black hole ``radiates" as a perfect black body; its
physical temperature is $\kappa/ 2 \pi$.

\section{Some Open Issues}

In this section, I will briefly discuss several open issues in quantum field
theory in curved spacetime. My purpose in doing this is
merely to give the reader
a flavor of some topics of current research interest. No attempt will be
made to provide
a comprehensive account of all present research in quantum field
theory in curved spacetime.

One key issue that remains unresolved concerns the treatment of
back-reaction. There is general
agreement that, in the context of the semiclassical approximation,
back-reaction effects should be described by the semiclassical
Einstein equation (\ref{see}). However, the following
two significant difficulties of
principle arise when one attempts to solve this equation and extract
physically relevant solutions: (i) As mentioned above, there is a local
curvature ambiguity in the definition of $<T_{ab}>$. By requiring
that the ambiguous, conserved
local curvature terms have the ``correct dimension", this
ambiguity for $4$-dimensional
spacetimes can be reduced to a two parameter family. However,
as discussed in \cite{w1}, at least one of these two parameters
cannot, in principle, be determined by arguments
involving only quantum field theory in curved spacetime (as opposed
to quantum gravity). Thus, there is a fundamental ambiguity in eq.
(\ref{see}). (ii) Equation (\ref{see}) is of a ``higher derivative" character
than the classical Einstein equation, and it admits many -- presumably
spurious -- ``run-away" solutions. However, it is not clear, in general,
how to distinguish between the ``physical" and the ``unphysical,
spurious" solutions.
Some proposals in this regard have been given by Simon \cite{sim}, and
further discussion can be found in \cite{fw}.

It is worth noting that the mathematical
situation with regard to treating back-reaction is considerably improved
when one considers $2$-dimensional ``dilaton gravity" models
\cite{cghs}, \cite{rst}.
In particular, there now are no conserved local
curvature terms of the correct dimension, so difficulty (i) does not
arise. Difficulty (ii) also is considerably alleviated. In addition,
in the conformal vacuum state $<T_{ab}>$
is given by a relatively simple, local expression in the conformal factor,
so the back reaction equations are considerably more tractable than in
the $4$-dimensional case.

The second open issue I wish to mention concerns energy conditions
which may hold for $<T_{ab}>$. Even for a Klein-Gordon field
in Minkowski spacetime, it is easy
to find quantum states for which $<T_{ab}>$ violates any of the
local positive energy conditions which hold for classical fields. However,
it is possible that $<T_{ab}>$ may still satisfy some nontrivial {\it global}
positive energy conditions. The most interesting of these conditions is the
``averaged null energy condition" (ANEC), which states that for any
quantum state and for any complete null geodesic, we have
\begin{equation}
\int <T_{ab}> k^a k^b d \lambda \geq 0
\label{anec}
\end{equation}
where $\lambda$ denotes the affine parameter of the geodesic and
$k^a$ denotes its tangent. The validity of ANEC is sufficient for proofs of
the positive energy theorem \cite{psw} and for ``topological censorship"
\cite{fsw}, so many of the key results of classical general relativity
continue to hold when the pointwise energy conditions are replaced by
ANEC. Although ANEC holds for Minkowski spacetime \cite{k},
\cite{wy}, it is known to fail
for arbitrary curved spacetimes (in $4$-dimensions)
\cite{wy}. Nevertheless, some recent research has indicated that
(i) ANEC may come ``close enough" to holding to exclude wormholes with
curvature everywhere much smaller than the Planck scale \cite{fr}
and (ii) when the semiclassical Einstein equation (\ref{see}) is imposed,
a version of ANEC may hold wherein one
``transversely averages" (with a suitable smearing function) over null
geodesics within
roughly a Planck length of the given geodesic \cite{fw}. Thus,
it is possible that violations of ANEC may be confined to regimes where
the semiclassical approximation is not applicable.

The final topic of present research which I wish to mention concerns
quantum field theory on spacetimes with closed causal curves. Consider,
for definiteness, a causality
violating spacetime which is ``globally hyperbolic outside of a
compact set", i.e., a spacetime obtained by modifying the metric of a
globally hyperbolic spacetime in a compact region so as to produce closed
timelike curves within that region. It seems plausible that -- at least
within a suitable subclass of such spacetimes -- one will have a well
defined, deterministic dynamics for classical fields \cite{fm}. Does there
exist a similarly well defined quantum field theory on such spacetimes?

For linear quantum fields, one needs only the symplectic vector space
structure of the classical solution space in order to construct
the Weyl algebra, so if the classical
dynamics is well behaved there should be no
difficulty defining fundamental observables and states. In particular, it
should be possible to obtain
a well defined, unitary $S$-matrix describing scattering processes.
However, the association of a solution, $Ef$, to
every test function $f$ (see eq.(\ref{id})) uses global hyperbolicity, so
it is not clear that the field observables can
still be interpreted as distributions
on spacetime. Even if they can, it can be shown that the condition of
``F-compatibility" \cite{kay} must fail \cite{krw}
at least at some points of the chronology
horizon, so the local behavior of the field observables must be different
from that occurring in globally hyperbolic spacetimes. Furthermore, the
local Hadamard condition cannot be satisfied everywhere on the
chronology horizon \cite{krw}. Thus, although in certain examples it is
possible to choose states where the stress-energy tensor remains finite
as one approaches the chronology horizon
\cite{kra}, the stress-tensor always must
be singular (or ill defined) at least at some points of the chronology
horizon. Thus, it seems far from clear that the quantum theory of linear
fields can be sensibly defined even on the class of causality violating
spacetimes which are ``globally hyperbolic outside of a compact set".

The situation for nonlinear fields appears to be considerably worse in that
conventional perturbation theory yields an $S$-matrix which is
non-unitary (in the sense of not conserving probability) \cite{fps}.
Alternative ideas for constructing quantum field theory on causality
violating spacetimes are currently being actively pursued by a number of
authors \cite{p}-\cite{haw2}.

\section{Summary}

The theory of linear quantum fields propagating on a fixed, globally
hyperbolic, curved spacetime is completely well posed mathematically
insofar as the definition of the fundamental observables
(or, equivalently, the smeared field operators) are concerned,
and $<T_{ab}>$ is well defined (for Hadamard states) up to local
curvature term ambiguities.
This provides the necessary tools to investigate many
phenomena of interest involving quantum effects occurring in strong
gravitational fields. Nevertheless, some interesting and important
issues remain open, such as the ones mentioned in the previous
section.

One may hope that some of the insights obtained from the study of
quantum field theory in curved spacetime -- particularly, the close
relationship between Killing horizons and thermal states as seen in the
Unruh and Hawking effects -- will be helpful in guiding the development
of a quantum theory of gravity.

{\bf Acknowledgment}

This research was supported in part by
National Science Foundation
grant PHY-9220644 to the University of Chicago.

\end{document}